# MLDAS: Machine Learning Dynamic Algorithm Selection for Software-Defined Networking Security

Pablo Benlloch[1], Oscar Romero[1], Antonio Leon[1] and Jaime Lloret[1,*]

[1] Departamento de Comunicaciones, Universitat Politecnica de Valencia; pablo.benlloch-caballero@uws.ac.uk, oromero@dcom.upv.es, aleon@dcom.upv.es, jlloret@dcom.upv.es
* Correspondence: jlloret@dcom.upv.es

**Abstract**

Network security is a critical concern in the digital landscape of today, with users demanding secure browsing experiences and protection of their personal data. This study explores the dynamic integration of Machine Learning (ML) algorithms with Software-Defined Networking (SDN) controllers to enhance network security through adaptive decision mechanisms. The proposed approach enables the system to dynamically choose the most suitable ML algorithm based on the characteristics of the observed network traffic. This work examines the role of Intrusion Detection Systems (IDS) as a fundamental component of secure communication networks and discusses the limitations of SDN-based attack detection mechanisms. The proposed framework uses adaptive model selection to maintain reliable intrusion detection under varying network conditions. The study highlights the importance of analyzing traffic-type-based metrics to define effective classification rules and enhance the performance of ML models. Additionally, it addresses the risks of overfitting and underfitting, underscoring the critical role of hyperparameter tuning in optimizing model accuracy and generalization. The central contribution of this work is an automated mechanism that adaptively selects the most suitable ML algorithm according to real-time network conditions, prioritizing detection robustness and operational feasibility within SDN environments.

## 1. Introduction

Network security has always been one of the fundamental components of communication networks. Users increasingly demand secure networks, where they can browse freely knowing that their personal data will be safe from intrusions and potential impersonations. An Intrusion Detection System (IDS) is one of the elements that configure the security of a network, and its objective is to constantly monitor the events that occur in the network system and identify possible attacks. Machine Learning (ML) makes its appearance in anomaly-based IDS, where models are trained to identify activities considered as normal and intrusions. Thus, supervised learning algorithms are used for intrusion detection [1].

The specific characteristics of Software-Defined Networking (SDN) controllers (centralized control logic, global network visibility, software-based traffic analysis, and dynamic updating of routing rules and tables) improve the security of the network when combined with ML-based intrusion detection algorithms. However, some SDN-based attack detection mechanisms have several limitations. This may be because they were primarily designed for traditional architectures, and the abstraction of the control logic has necessitated a rethinking of the model. Two of the limitations they present are the Misbehavior Attack and the NewFlow Attack. These are studied in [2], which proposes a secure system that periodically collects statistics from forwarding elements and applies ML classification algorithms. They claim that the proposed solution generates a more secure SDN architecture capable of intelligently adapting to network conditions and reacting to changes.

SDN-based attack detection mechanisms are focused on traffic inspection. Usually, this inspection is only performed on the first packet of the traffic flow, and the same decision is applied to the remaining packets in the flow. Therefore, if an attacker generates the first packet with normal behavior and the rest with anomalies, it can lead to a Misbehavior Attack. On the other hand, SDN-based environments that include a switch unable to locate the flow to which a packet belongs will send the packet to the controller to receive instructions. Attackers can exploit this characteristic of the SDN paradigm to generate a large number of unlabeled flows to SDN switches. As a result, it can lead to a DDoS attack (Distributed Denial of Service) [3].

In this work we propose a lightweight ML model to detect DDoS attacks in real-time in SDN networks. It uses a dataset gathered from an emulated network infrastructure using Mininet and real network traffic, enhancing the reliability and applicability of the findings. This model, implemented in the Ryu Controller, is lightweight, minimizing additional computational overhead. The controller notifies the necessary information (source IP, destination IP, source port, destination port, protocol type, etc.) for making a decision. Based on the obtained result, the controller will finally update the SDN switches with new rules, as well as update the trained ML model. Additionally, most publications use pre-made or open datasets. However, in this proposal, our dataset has been obtained in an emulated infrastructure, and the attacks launched are real, as real tools such as hping3 have been used to execute the attacks. Therefore, the traffic morphology is completely real, and the attack behavior has not been simulated. The final results demonstrate high accuracy and performance, showcasing its effectiveness in detecting and classifying network anomalies in SDN environments.

The rest of the paper is structured as follows: Section 2 provides an overview of related work in the field of network security and machine learning. Section 3 describes the methodology used in this study, including the dataset, metrics analysis, and machine learning models. Section 4 presents the experimental results and their analysis. Finally, Section 5 concludes the paper and discusses future work.

## 2. Related Works

In this section we are going to review the existing literature on the integration of machine learning and software-defined networking for network security. We will discuss the limitations of traditional network security mechanisms and how machine learning can be used to enhance them. We will also explore the different approaches and techniques used in previous studies and their effectiveness.

In recent years, numerous studies have been conducted on network security and DDoS attack detection using ML. Many studies have analyzed the importance of network security. In [4], specific types of security problems and several effective countermeasures are discussed, while in [5], the connotation of network security, the research content of network security, the framework of network security information, and the dynamic model of network security are summarized. Network security problems are studied from four aspects: network hardware, network software, network managers and network users. A network security study from the perspective of the data value chain is analyzed in [6], where the whole process is composed of five successive stages: factor acquisition, model representation, measurement establishment, solution analysis, and situation prediction. Authors in [7] discuss the basics of network security, emphasizing the distinction between the network and transport layers in Internet traffic, and highlighting that the Transmission Control Protocol is essential for reliable data delivery.

Different reference models have been proposed in previous works, such as in [8]. This work proposes a network security reference model, consisting of four layers from the point of view of network configuration elements: inter networks, single network, equipment, and data.

Other studies examine network technologies. In [9], network security technologies are studied in detail: authentication, data encryption technology, firewall technology, intrusion detection system (IDS), antivirus technology and virtual private network (VPN). Authors in [10] present a general analysis of security technology, internet vulnerabilities, and attack methods through the internet. Best practices on the topics of authentication, authorization, auditing, firewall, intrusion detection & monitoring, and prevention are provided in [11].

Several specific studies focus on DDoS attacks in SDN networks. For instance, in [3], authors identify and defend DDoS attacks based on machine learning for the campus network. The method is divided into 3 parts: traffic collection module, DDoS attack identification module and flow table delivery module. The Support Vector Machine (SVM) is applied to identify the DDoS traffic and Ryu Controller is employed to build the flow table decision delivery module. HTTP flood attacks against web servers are studied, considering eight data packet features. In [12], an intrusion detection system technique is proposed to detect some DDoS attacks on an SDN environment with the implementation of four different machine learning: Multiple Layer Perceptron (MLP), Support Vector Machine (SVM), Decision Tree, and Random Forest algorithm.

The Scapy tool is used to generate the attacks and a Mininet with POX Controller to simulate the SDN environment. The Random Forest algorithm and the Decision Tree algorithm have good results, while MLVP and SVM do not provide the best results. In [13], machine learning and statistical methods are combined to improve the detection of DDoS attacks in SDN networks. The proposed method consists of three sections: three collector, entropy-based and classification. Floodlight Controller is used for collecting flows, and UNB- ISCX and CTU-13 datasets are selected. Acceptable results are obtained, but the method increases CPU load.

Some studies evaluate several machine learning techniques, like [14], using a Ryu Controller, and [15], focused for low rate. A hybrid machine learning model to protect the controller from DDoS attacks is proposed in [16], combining two machine learning based models, SVM and Self Organized Map. The controller is POX, and packet generation is done with Scapy tool. The results show an accuracy of 98%.

Recent studies employing deep learning techniques, such as convolutional (CNN), recurrent (RNN), and gated recurrent unit (GRU) architectures, have demonstrated high accuracy in DDoS and intrusion detection tasks [17–20]. However, these approaches generally involve higher computational complexity and are optimized for offline or centralized analysis, which constrains their real-time applicability within SDN controllers. In contrast, the proposed MLDAS framework emphasizes real-time adaptability and efficient model switching directly within the SDN controller, achieving a favorable trade-off between detection accuracy and operational latency.In [17], authors propose a method based on the analysis of single IP flow records, which uses the Gated Recurrent Units deep learning method to detect DDoS and intrusion attacks. It uses public datasets and provides faster responses, but this model uses a complex framework.

There are many data packet features to be studied with ML. A different number and type of these features are used in the different works. In [1], several feature selection methods for machine learning on DDoS detection are evaluated, including a comparative analysis of feature selection and machine learning classifiers. Good results are obtained by Random Forest algorithm.

Most of the works are focused on DDoS attacks detection in standard SDN network, but in some works, the scope of study is expanded, like in [18], where A near real-time SDN security system that prevents DDoS attacks in IoT networks is studied. Good results are obtained but accuracy decreases after some time. Also, in [19] it is proposed a DDoS attack detection method for 5G and B5G. It has good results, but it has a complicated structure that may result in lower performance in real-time networks. Finally, authors in [20] use a Deep Neural Network to detect DDoS attacks. They get high accuracy, but it decreases for multicast classification.

In [21], DDoS attacks from the traces of the traffic flow are discussed. Different machine learning algorithms are used, such as Naive Bayes, K-Nearest neighbor, K-means, and K-medoids to classify the traffic as normal or abnormal. Authors in [22] provide a systematic benchmarking analysis of the existing machine learning techniques for the detection of malicious traffic in SDNs. They identify some limitations in theses classical machine learning based methods, and lay the foundation for a more robust framework. Regarding defense mechanisms to mitigate DoS attacks in an SDN network environment, some applications are reviewed in [23].

In [24], the authors propose a method based on machine learning to detect Denial of Service (DoS) attack in data plane devices, i.e., the OpenFlow switches, resulting from flow-table overflow. They create SDN dataset using Mininet and features are extracted from switch-controller communication. Several tools for threat detection have been studied, for example, [25] and [26], where is introduced the principle of penetration testing for network intrusion detection, and Penetration Testing Execution Standard (PTES) tool is used, respectively.

In the field of IoT, studies on security have been conducted IoT security. In [27], the authors present a machine learning-based approach to detect DDoS attacks in an SDN-WISE IoT Controller. The machine learning DDoS detection module is integrated into the SDN-WISE Controller and uses Naive Bayes (NB), Decision Tree (DT), and Support Vector Machine (SVM) algorithms to classify SDN-IoT network packets. The authors in [28] propose a new network intrusion detection method that is appropriate for an Internet of Things network. The proposed method is based on a conditional variational autoencoder with a specific architecture that integrates the intrusion labels inside the decoder layers. The method can perform feature reconstruction, that is, it is able to recover missing features from incomplete training datasets. In [29], two dimensions of security assessment are explored, using vulnerability information of IoT devices and their underlying components and SIEM logs captured from the communications and operations of such devices in a network to propose the notion of an attack circuit. These measures are used to evaluate the security of IoT devices and the overall.

In contrast to previous works that rely on static models or pre-trained classifiers, this paper proposes a novel adaptive framework that dynamically selects the most suitable ML algorithm based on real-time traffic conditions in SDN networks. Unlike prior approaches that often depend on public or synthetic datasets, our model is trained and validated using traffic generated in both emulated environments, via Mininet, and a controlled physical testbed composed of interconnected PCs, simulating realistic attack scenarios. Additionally, the ML model is directly integrated into the SDN Controller Ryu, enabling live packet monitoring, real-time predictions, and immediate update of flow rules. This combination of dynamic algorithm selection, real traffic training, and direct controller integration represents a significant step forward compared to existing methods, which typically focus on accuracy in offline datasets without addressing performance, adaptability, or deployment practicality in live SDN environments.

## 3. System Proposal

In this study, the assumed threat model considers an external adversary capable of launching volumetric or protocol-based DDoS attacks against SDN-controlled networks. The attacker can manipulate packet rates, ports, and source IP addresses (including spoofing) to disrupt legitimate services, but is not assumed to have access to or control over the SDN controller, forwarding devices, or host operating systems. Accordingly, the proposed defense mechanism focuses on flow-level detection and mitigation of external traffic anomalies rather than internal compromise.

3.1 *System Architecture*

The target of this work is the traffic classification using ML in SDN networks, two using a dynamic algorithm to select the best machine learning model. The classification should differentiate between legitimate traffic and potential DDoS attacks. To achieve this objective, the design and development of a network emulation environment have been proposed to generate random traffic, collect the data, and create a dataset that allows training different models, and finally the Machine Learning Dynamic Algorithm Selection for Software-Defined Networking Security (MLDAS) module will select the ML model to fulfill the final function of traffic classification.

The proposed method uses a three-layer architecture, shown in Figure 1.

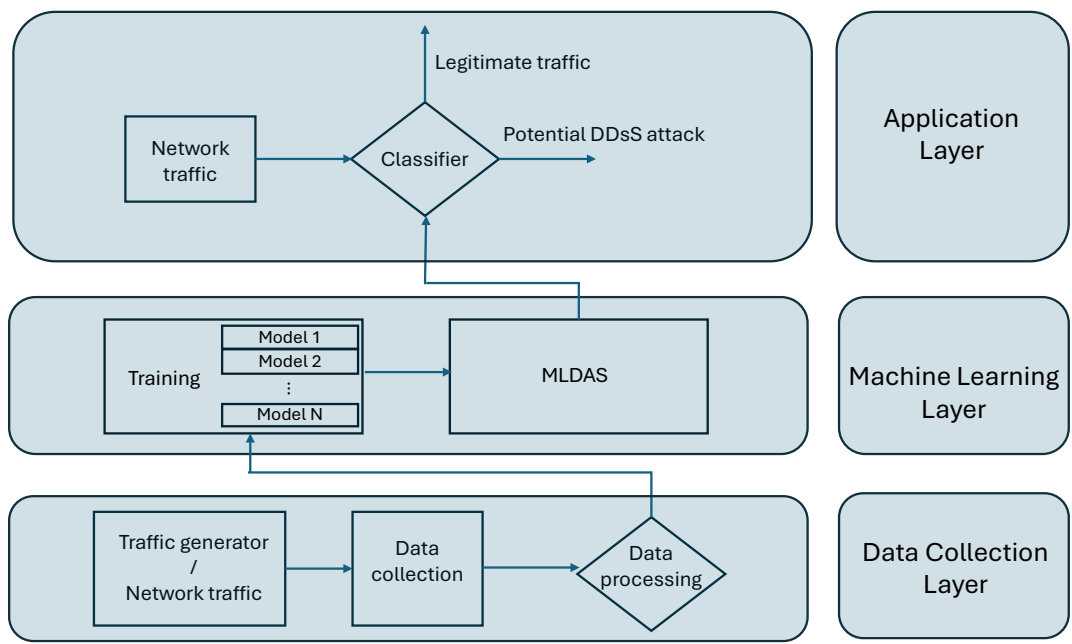

**Figure 1.** Traffic classification architecture.

- Layer 1: Data Collection Layer.

The proposed system can be implemented in both real networks and network emulation environments created using Mininet, as the developed software is compatible with both scenarios. In this layer, either captured network traffic or a traffic generator can be used. In the case of a traffic generator, random traffic is initially produced by a Python script running on multiple network hosts, whether physical or emulated. A hybrid network configuration is also supported, where a Mininet-emulated environment is connected to a physical network. The resulting data is then collected by the controller, which can be either an emulated controller running within Mininet or an external physical controller. A physical controller is generally preferred due to its greater computational capacity, especially in large-scale networks with high traffic volumes. Layer 1 is thus focused on the collection of flow-level statistics such as packet counts, byte counts, and duration, which are exported to the upper layers for further analysis.

- Layer 2: Machine Learning Layer.

In this layer, the preprocessed data is used to train and test various machine learning models. Multiple models are evaluated using different combinations of metrics. All results are forwarded to the Machine Learning Dynamic Algorithm Selection module, where the metrics are analyzed to compute the accuracy and performance of each model. Based on this analysis, the module selects the best-performing model to be used by the traffic classification component. It is important to note that this analysis is not performed only once. Depending on network conditions and traffic patterns, new data is continuously collected and reprocessed, and the machine learning models are re-evaluated using updated combinations of metrics.

- Layer 3: Application Layer

In this layer, the selected machine learning model is used to classify network traffic as either legitimate or potentially associated with DDoS attacks. The model is integrated with the SDN controller to enhance network security. Depending on the characteristics of the network traffic, the machine learning model, selected by the previous layer, may change dynamically to ensure optimal performance.

*3.2 Feature Extraction*

The statistics collected at Layer 1 are transformed into a set of flow features used for training and prediction. The selected features are lightweight to compute at the controller while providing enough discriminative power to distinguish between legitimate and malicious behaviors. Specifically, the feature set includes: flow duration, number of packets per flow, total bytes per flow, average packet size, inter-arrival time statistics (mean, variance), TCP flag counts (SYN, ACK, FIN), and flow directionality (ratio of incoming/outgoing packets). These features capture complementary aspects of traffic: volume, temporal dynamics, and protocol semantics. For example, DDoS floods exhibit very high packet counts with short inter-arrival times, port scanning generates numerous short-lived flows with distinctive flag patterns, and brute-force attempts manifest as repeated connections to the same service port. This feature set therefore provides a concise yet discriminative representation of traffic behavior while keeping the collection overhead at the SDN controller low. The extracted features serve as input for the MLDAS decision process, which is described in the following subsection. It is important to note that the discriminative power of these features may vary under adversarial or camouflage-based attacks, where malicious traffic attempts to mimic legitimate flow characteristics. Evaluating the robustness of the selected features under such adaptive conditions represents an essential next step to ensure stability and reliability in real-world deployments.

The selected flow features were chosen based on the characteristics of SDN traffic and typical DDoS attack behaviors. Metrics such as packet counts, byte totals, and flow duration capture volumetric characteristics of the traffic, while inter-arrival statistics reflect temporal dynamics. Additionally, TCP flag distributions and directionality ratios encode protocol-level semantics that help distinguish legitimate sessions from flooding or scanning activities. These features were preferred because they are readily available through OpenFlow statistics and can be computed at the controller with minimal overhead, ensuring both discriminative power and real-time feasibility.

Preliminary correlation checks were conducted to verify that no strong redundancy existed among the selected metrics. However, a more systematic dimensionality reduction and feature selection study (e.g., PCA or regularization-based methods) will be pursued in future work to refine model efficiency and complexity.

### 3.3 MLDAS Decision Process

The MLDAS module applies a reproducible two–stage decision process to select the optimal model. The complete decision logic of the MLDAS module is summarized in Algorithm 1 and graphically represented in Figures 2 and 3, which describe in detail the evaluation steps, thresholds, and switching conditions required to replicate the procedure. First, all candidate models (Decision Tree Classifier, Decision Tree Regressor, Random Forest Classifier, and Linear Regression) are evaluated against a rolling RMSE threshold ($S_{min}$). Models exceeding this threshold are discarded. Second, among the valid candidates, the model with the lowest prediction latency is selected, using training latency as a tie-breaker. The evaluation is triggered either periodically every W flows or on demand when the rolling error of the active model increases beyond a tolerance ($\varepsilon\_err$) or its rolling accuracy falls below $A\_min$. To avoid oscillations between models, hysteresis is enforced through a minimum dwell time ($T\_dwell$) before switching and a minimum relative improvement threshold ($\tau\_switch$). The procedure is summarized in Algorithm 1.

In practice, the MLDAS module performs a re-evaluation every W = 200 flows or whenever the rolling RMSE of the active model increases by more than $\varepsilon\_err$ = 0.2 % or its accuracy falls below A_min = 98 %. Candidate models exceeding the Smin threshold are discarded, and a switch is authorised only if the improvement of the new model exceeds $\tau\_switch$ = 5 % after a minimum dwell-time of T_dwell = 600 flows. These parameters collectively ensure adaptive yet stable operation, preventing oscillations between models while maintaining responsiveness to performance degradation.

**Algorithm 1 – MLDAS: Dynamic Model Selection**

Inputs:
```
  C = {M1, M2, ..., Mk}       // Set of candidate models
  Smin                        // Maximum acceptable RMSE threshold
  W = (#flows per evaluation) // Window size for re-evaluation (e.g., 200)
  T_dwell                     // Minimum dwell time before allowing a switch (e.g., 3W)
  ε_err                       // Error degradation tolerance (e.g., +0.2% absolute)
  A_min                       // Minimum acceptable rolling accuracy (e.g., 98%)
  τ_switch                    // Minimum improvement required to switch (e.g., 5%)
  Mode = {periodic | event}   // Re-evaluation triggers (periodic or degradation-based)
```

Initialization:
```
 For each Mi ∈ C:
   Evaluate RMSEi on validation data
   Measure train_time_i and pred_time_i (in the target environment)
 C* ← {Mi ∈ C | RMSEi ≤ Smin}
 M_current ← argmin_{Mi ∈ C*} (pred_time_i, tie-break: train_time_i)
 last_switch ← 0
```

Online operation (for each incoming flow batch B of size W):
```
 1) Use M_current to predict B and apply SDN actions if required.
 2) Compute RMSE_roll(B) and Acc_roll(B) for M_current (rolling).
 3) If Mode = periodic and (flows_processed mod W = 0) OR
       Mode = event and (RMSE_roll(B) > RMSE_current + ε_err OR Acc_roll(B) < A_min):
     a) For each Mi ∈ C:
```

     Estimate RMSEi_roll on B (or in a mirrored buffer) and measure pred_time_i
     C† ← {Mi | RMSEi_roll ≤ Smin}
     If C† = ∅: keep M_current and continue
     M_best ← argmin_{Mi ∈ C†} (pred_time_i, tie-break: train_time_i)
  b) Hysteresis and anti-flapping:
     If (flows_processed - last_switch) < T_dwell: keep M_current
     Else if Improvement(M_best vs M_current) < τ_switch: keep M_current
     Else: M_current ← M_best; last_switch ← flows_processed
4) Proceed with the next batch of flows (loop continues)

The periodic evaluation of model performance also provides a mechanism for implicit adaptation to gradual concept drift, since variations in traffic behavior are captured through the continuous monitoring of error and accuracy thresholds. Explicit drift-detection or incremental-learning strategies will be considered in future work to further enhance the framework's long-term adaptability.

The same decision process is illustrated in Figure 2 as a flowchart. This graphical representation highlights the sequential evaluation of candidate models, the quality filtering based on RMSE, the selection of the model with the lowest prediction latency, and the application of hysteresis conditions.

| Parameter | Description | Value used |
|---|---|---|
| $S_{min}$ | Maximum acceptable RMSE for candidate models | From validation (Sec. 4.4) |
| $W$ | Re-evaluation window (number of flows) | 200 flows |
| $A\_min$ | Minimum acceptable rolling accuracy | 98% |
| $\varepsilon\_err$ | Error degradation tolerance | 0.2% |
| $T\_dwell$ | Minimum dwell time before switching | 600 flows |
| $\tau\_switch$ | Minimum relative improvement to allow switching | 5% |

**Table 1.** Operational thresholds and re-evaluation parameters of the MLDAS model-selection mechanism.

3.3.1. Operational Thresholds and Model Re-evaluation Policy

Table 1 summarizes the operational thresholds and the parameters used for dynamic model selection in MLDAS. The minimum acceptable error threshold, *Smin*, is defined as an upper bound on acceptable prediction error, derived from the validation RMSE distribution of the candidate models to ensure reliability while avoiding overly restrictive filtering. Model evaluation is performed either periodically or in an event driven manner, as described below.

Model evaluation is triggered either periodically every *W* flows or in an event driven manner when the rolling RMSE of the active model increases beyond *ε_err* or its rolling accuracy falls below *A_min*. To ensure stability and avoid frequent changes between models, a hysteresis mechanism is applied using a minimum dwell time (*T_dwell*) and a minimum relative improvement threshold (*τ_switch*). This design ensures that a model change occurs only when a sustained and meaningful performance improvement is observed.

The parameters of the selection policy were set following common practices in adaptive learning systems and validated through preliminary experiments. A re-evaluation window of

W=200 flows provides a compromise between responsiveness and stability, as smaller windows increase noise while larger ones delay adaptation.

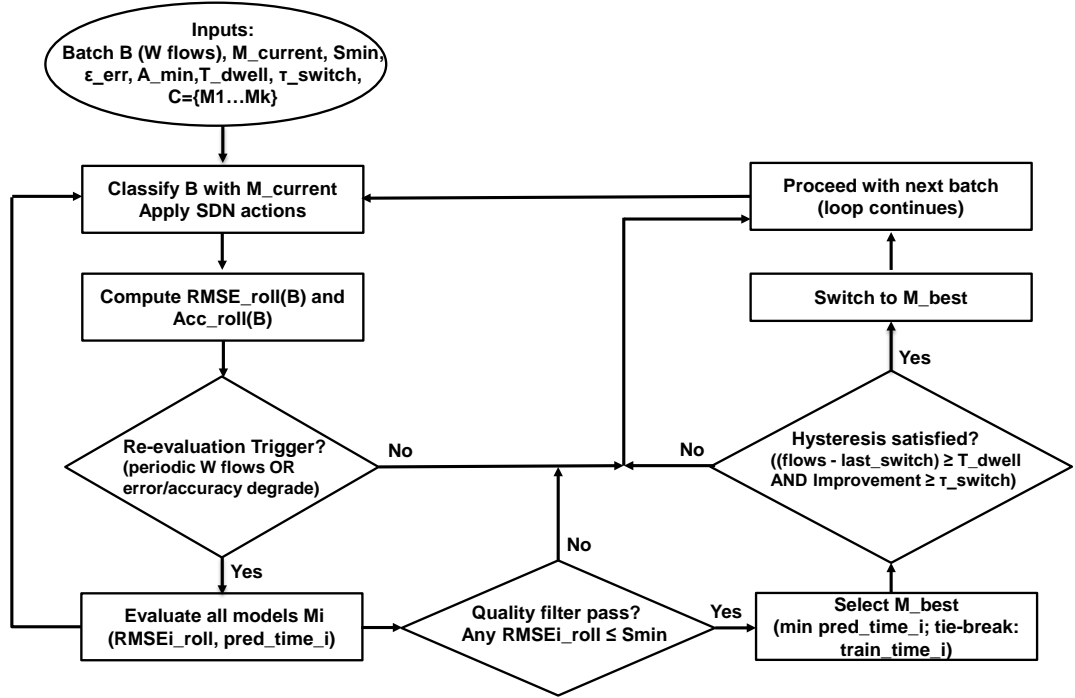

**Figure 2.** MLDAS dynamic model selection process.

The error tolerance $\varepsilon\_err$=0.2% was chosen to detect meaningful degradations without reacting to random fluctuations. A minimum accuracy threshold $A\_min$=98% was derived from the cross-validation results in Section 4.6, where this value consistently separated normal and attack traffic. Hysteresis parameters $T\_dwell$=3W and $\tau\_switch$=5% are standard safeguards in control systems, preventing frequent oscillations and enforcing that only substantial improvements trigger a model switch. These settings ensured that the MLDAS module remained both effective and stable across all tested scenarios.

In addition to the global decision process shown in Algorithm 1, Figure 3 provides a detailed view of the RMSE-based quality filtering step performed within the MLDAS selection process. The inputs are *hi(x)*, the predictions generated from the dataset containing the metrics to be predicted using machine learning model i, and *yi(x)*, the reference label used during the validation phase. Accuracy is evaluated using the Root Mean Square Error formula. This error value is stored in variables *RMSEi*, one for each model i. Additionally, training and prediction times are calculated, as the goal is not only to find the most accurate algorithm but also one with good performance in order to minimize hardware requirements. Once all *RMSEi* values are obtained, they are compared against a minimum required threshold, *Smin*, to ensure the results are sufficiently reliable. The value of *Smin* should be equal to or lower than the maximum of all *RMSEi* values. The closer *Smin* is to the maximum *RMSEi*, the more restrictive the selection of candidate machine learning algorithms for the next stage will be. While it is also possible to sort the *RMSEi* values from highest to lowest and, for instance, select the top three models, this may be arbitrary. It is more effective to rely directly on the numerical *RMSEi* values. The selected algorithms are renumbered as j, k, ..., p. In the next stage, the training and prediction times of these selected models are compared to determine the one with the lowest value, which we identify with the subscript z. Thus, the machine learning model chosen for the next stage is model z, which will be used to classify the traffic. The MLDAS module is both flexible and iterative, continuously performing calculations based on newly collected traffic to determine whether the current model remains optimal for classification or whether a different model is better suited under the updated conditions. The complete decision logic and threshold interactions are summarized in Algorithm 1 and illustrated in Figures 2 and 3.

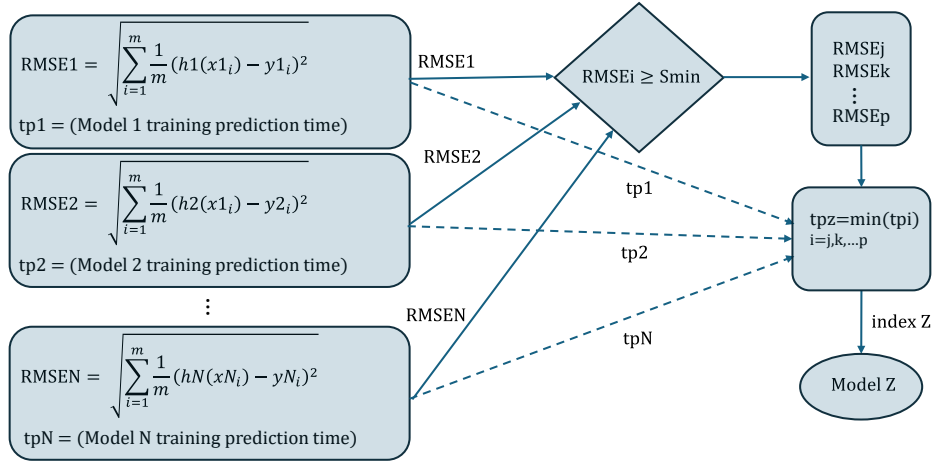

**Figure 3.** RMSE-based evaluation within the dynamic model selection process.

*3.4. Summary and Integration*

To ensure the model remains optimal over time, the MLDAS module performs this evaluation periodically. The reevaluation is triggered after a fixed number of classified flows, or when the error rate of the currently selected model exceeds a predefined tolerance. This trigger mechanism ensures that the system adapts dynamically to evolving traffic patterns while avoiding excessive switching between models. By combining both the accuracy of the model, measured via *RMSEi* and compared against a minimum acceptable threshold *Smin*, and the performance metrics, the system balances detection quality with real-time operational constraints. In this context, real-time refers to the ability of the system to analyze traffic directly as it is observed by the SDN controller, without relying on offline processing or delayed batch analysis. The prediction and decision-making are performed quickly, allowing timely responses to emerging threats and enabling live adjustment of flow rules.

Upon detecting an attack, the system updates the flow table of the corresponding SDN switch by installing a high-priority DROP rule targeting the source IP address of the malicious flow. This effectively blocks further traffic from the attacker without affecting other users. The rule is limited in scope to ensure that only the identified source is blocked.

The proposed framework combines three main components: (i) a data collection layer that gathers flow-level statistics from the SDN controller, (ii) a feature extraction stage that derives lightweight but discriminative attributes from the collected statistics, and (iii) a dynamic decision process that automatically selects the most suitable ML model according to current network conditions. Together, these components enable the SDN controller to perform accurate and efficient intrusion detection in real-time, while adapting to traffic variability.

## 4. Simulation and Physical Network Environments

*4.1. Network Diagram*

The experimental evaluation was implemented in both emulated and physical SDN testbeds. In the emulated setup, Mininet (v2.3.0) and Open vSwitch (v2.17) ran on a dedicated server (Intel Core i7 CPU, 32 GB RAM, Ubuntu 22.04, Linux kernel 5.15), with the Ryu Controller (v4.34) configured with OpenFlow 1.3. In the physical testbed, the same server hosted the Ryu Controller, while traffic was generated and processed by multiple hosts equipped with Intel Core i7 CPUs and 16 GB RAM, interconnected through HP ProCurve-3500 OpenFlow-enabled switches. An additional server was available in the testbed to support traffic monitoring and dataset collection.

This dual-testbed design enabled evaluation under both reproducible emulation and realistic physical conditions. The network diagram shown in Figure 4 was used consistently across both environments, ensuring comparable behavior in both setups.

In order to evaluate the proposal, two datasets were used: one captured in a real SDN deployment and one generated in an emulated environment with the same topology (5 traffic PCs, 1 server, and the Ryu Controller). In the real deployment, we collected approximately 7 hours of traffic, yielding about 120k flows and 15 million packets. Legitimate traffic comprised web browsing and HTTPS downloads, DNS queries, SSH sessions, file transfers (FTP/SCP), and video streaming; background load also included synthetic TCP/UDP flows (iperf). Attack traffic included TCP SYN floods, UDP floods, port scanning (nmap), and SSH brute force (hydra). In the emulated testbed we generated a comparable volume, around 6 hours of traffic with 100k flows and 12 million packets, using the same mix of legitimate applications and attack types.

To generate and control traffic across the test environments, custom Python scripts were developed for both scenarios. The traffic was categorized into two types, legitimate and DDoS attack traffic, and labeled accordingly to enable the use of supervised learning models, as described in later sections. This classification is essential for training and validating the proposed machine learning-based detection framework.

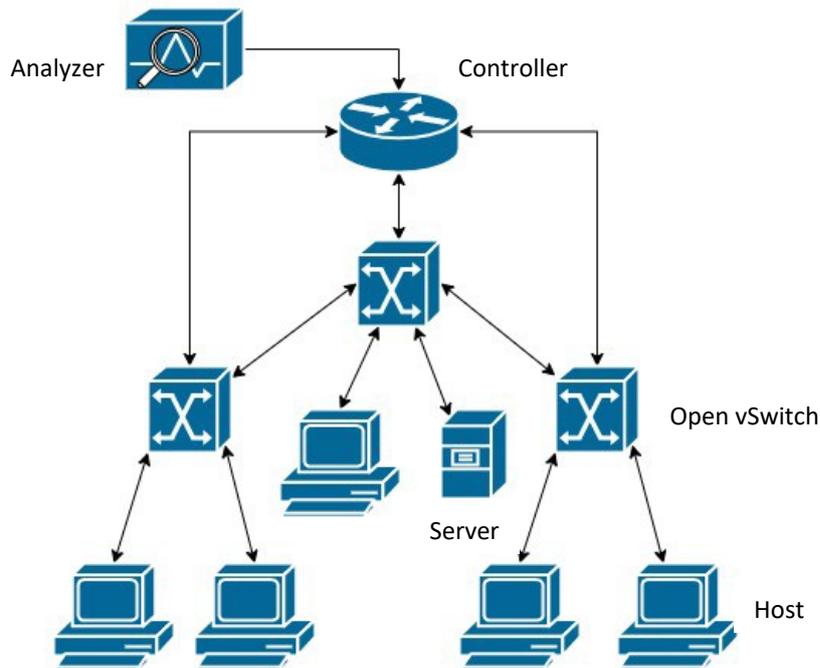

**Figure 4.** Network topology.

The traffic generated by the script to represent legitimate network activity was produced over multiple iterations, ensuring repeatability and variability. Each iteration randomly assigned legitimate flows between different host pairs and varied protocol types, packet sizes, and inter-packet intervals. This process helped capture diverse traffic patterns, improving the generalization of the trained models.

1. A host is randomly chosen to act as the server.
2. The server checks if it has the necessary files to run the service. If not, it copies them from a backup folder.
3. The server sets up an HTTP and FTP web service.
4. The server starts an Iperf server tool that will be consumed by the rest of the hosts on ports 5050 for TCP and 5051 for UDP.
5. The following steps are repeated 50 times:
a) A host A is randomly chosen.
b) Another host B is randomly chosen, and a 100-packet exchange is created with host A using ICMP Ping at a rate of 0.2 ms per message.
c) Host A generates TCP and UDP traffic to the server's ports 5050 and 5051, respectively.
d) Host A consumes a web page from the server and verifies successful access.
e) Host A downloads a zip file from the server and verifies successful download.
6. The server is stopped, and the downloaded files are cleaned up.

On the other hand, the simulations performed by the script to generate attack traffic are carried out over several iterations, randomly selecting a victim host and an attacker host. To achieve this, the hping3 tool is used to launch four different types of DDoS attacks. The steps followed by the script can be outlined as follows:
1. A host is randomly chosen to act as the Server, which will be the victim of the simulation.
2. This server runs a web service, an FTP service, and Iperf for TCP and UDP as in the previous case.
3. A host is randomly chosen to act as the attacker.
4. The attacker initiates an ICMP Flood attack on the victim server.
5. The attacker initiates a UDP Flood attack on the victim.
6. The attacker initiates a TCP-SYN Flood attack on the servers.
7. The attacker launches a final LAND Flood attack on the server.

The different types of attacks consist of:
- ICMP Flood: It involves overwhelming the victim with ICMP echo request packets, preventing it from responding properly. In these simulations, the original source IP is hidden, which replaces the original IP with a random one.
- UDP Flood: It involves flooding the victim's UDP ports with IP packets containing UDP datagrams. The same techniques of IP hiding and packet size increase used in the previous ICMP Flood attack are employed.
- TCP-SYN Flood: In this attack, the server receives an overwhelming number of SYN requests to all its TCP ports, forcing it to respond with SYN-ACK to each request. However, the volume of requests saturates the server, causing it to stop providing the intended service. The same techniques of IP hiding and packet size increase used in the previous attacks are employed.
- LAND Flood: This Layer 4 attack is a variant of TCP-SYN. In this attack, the source IP and port are specified identical to the destination. As a result, when the SYN packets reach the server, it attempts to respond to itself in an infinite loop, eventually saturating the server as it consumes almost all of its CPU resources.

Labeling was based on the controlled execution schedule of the attacks: flows generated during attack periods were labeled as malicious, while all other flows were labeled as legitimate.

Per-flow labels were derived from start–end times and 5-tuple matching at the controller, ensuring consistency between the datasets.

Across all runs, the resulting datasets contained approximately 66% legitimate flows and 33% malicious flows. To mitigate potential class imbalance issues, training batches were constructed with balanced class proportions. For model development, we used a chronological split (70% training, 30% testing) to avoid look-ahead bias. In addition, 10-fold cross-validation was applied on the training set for hyperparameter tuning, ensuring reproducibility and providing a robust estimate of generalization performance. This dual setup allowed us to validate the framework in both an emulated environment and a real SDN testbed, supporting its practical applicability.

*4.2. Controller and Dataset*

In the case of the Mininet-based simulation network, the traffic generation script can be used in conjunction with the native controller provided by Mininet. However, this built-in controller offers limited flexibility and is not suitable for integrating custom logic or machine learning-based traffic analysis. For this reason, we also chose to use the Ryu Controller within the Mininet environment, allowing full programmability and consistency with the physical SDN testbed. Ryu is implemented in Python, which enables seamless integration with the traffic generation and ML components, so the entire system (Mininet, Ryu, and ML) can be developed using a single programming language.

First, we need to find a way to create the controller and connect it to the network topology. This can be done using a Python script that utilizes the functions specified in the Ryu documentation, which provide support for Layer 2 switches that form the Data Plane (DP) of the topology. We use the "packet_in_handler" method to redirect any packet arriving at any of the switch ports in our network. Using this function, the Ryu Controller listens to the switch port where a packet arrives and determines the best route to its destination, returning the outgoing port to the switch for packet forwarding. To store the route and ensure that future packets are sent through the same route, we program the controller using Ryu's "add_flow" function. Both of these functions, which control the DP from the CP (Control Plane), have been overridden compared to Ryu's native functions to include identification of the protocol used by the packet. This will later allow us to extend the generated dataset with additional information.

While public benchmark datasets such as CICIDS2017, NSL-KDD, or UNSW-NB15 are widely used in traditional network intrusion detection systems, they are not specifically designed for Software-Defined Networks. These datasets typically consist of pre-captured traffic traces intended for offline analysis, and therefore do not reflect the dynamic, real-time behavior characteristic of SDN environments. Moreover, they lack information related to flow rule updates, controller-switch interactions, and other SDN-specific events that are critical for validating an ML model integrated directly within the controller. For these reasons, this study relies on custom-generated traffic, both in emulated and physical SDN environments, to ensure that the model is evaluated under conditions that closely resemble actual SDN deployments, including real-time classification and response capabilities.

Although no backbone traffic was directly captured, the test scenarios were designed to emulate realistic conditions using both legitimate and malicious traffic patterns in controlled environments. While the exact statistical difference between generated and real-world traffic is not quantified, the use of two complementary testbeds and a dynamic model selection strategy enhances the ability of the system to adapt to diverse traffic behaviors, supporting its generalization capacity beyond the training context.

4.2.1. Metrics

One of the important decisions to be made is to select the type of metrics to be included in the dataset. This is considered very important since these data will define the behavior of the model when classifying traffic. First and foremost, we need to examine the set of traffic information provided by Ryu, which in turn relies on the data provided by the OF 1.3 protocol.

OF 1.3 offers two types of metrics that can help achieve the objective of this work. One type is port-based metrics, which provide information to the controller about the packet passing through the port and, if deemed an attack, take appropriate measures regarding that port. The second type refers to flow-based metrics, which consist of a set of destination IP, source IP, destination port, source port, and protocol used. The information provided by flow metrics is much more extensive than that obtained from port metrics, and they are not necessarily mutually exclusive. However, to streamline the dataset generation process, it is decided to only record metrics related to flows, and if necessary, the controller can request port-related information.

Furthermore, the actions to be taken once the traffic type is classified are also dependent on these metrics. These future actions could involve disabling the attacked port, for example, while allowing the rest of the network to continue functioning. This may be the initial idea for a novice engineer; however, it is not the most accurate approach, and flow-related information will help address the problem more effectively.

If the SDN system had a self-protection service based on port information, it would notice which ports are under attack and choose to cut communication with them, leaving the rest healthy and partially recovering the network. However, as mentioned earlier, this decision is not well thought out. The attacked port may host a critical communication service, and as a result, communication has been abruptly cut off not only by the attackers, but also by the autonomous self-protection system of the network. On the other hand, if the SDN system had a self-protection service based on flow information, and its actions were in line with that information, it could identify the number and nature of the flows causing the attacks and eliminate them immediately, thereby leaving the ports open for rescue communications and maintaining network stability.

4.2.2. Ryu Monitor

The traffic is monitored by Ryu Controller. Ryu will request the flows and packets passing through each of the switches, and when it detects the addition of new data frames, it will start writing them to the CSV. The metrics that will be recorded in the dataset can be summarized in the Table 2.

4.2.3. Running the Environments

Once the desired metrics to be recorded have been established and the controller has been programmed to write these metrics to a CSV file for the flows passing through the switches, the network activity, whether in the simulated Mininet environment or the physical testbed, generates the dataset. This data is subsequently analyzed using machine learning techniques, as described in the next section.

*4.3. Applying ML to the Dataset*

While deep learning techniques have gained popularity in recent years for network traffic analysis, this work focuses on traditional machine learning algorithms due to their lower computational cost and better suitability for real-time integration into the SDN controller. Since the classification runs directly inside the Ryu Controller, which operates in a constrained environment, efficiency is critical. Classical models such as Random Forest and SVM offer an optimal balance between accuracy and performance, and their interpretability further facilitates integration, debugging, and maintenance.

| Metric | Description |
|---|---|
| timestamp | timestamp associated with the moment when the data is recorded (ns) |
| datapath_id | ID of the OF switch that receives the packet |
| flow_id | ID formed by concatenating ip_src + tp_src + ip_dst + tp_dst + protocol |
| ip_src | source IP address of the flow |
| tp_src | source port of the flow. |
| ip_dst | destination IP address of the flow |
| tp_dst | destination port of the flow |
| ip_proto | IP protocol code used (icmp, tcp, udp) |
| icmp_code | icmp_code |
| icmp_type | icmp_type |
| flow_duration_sec | duration of the flow in s |
| flow_duration_ns | duration of the flow in ns |
| idle_timeout | pre-established flow idle timeout set to 20 |
| hard_timeout | maximum flow idle timeout, set to 100 |
| flags | TCP flags |
| packet_count | number of packets associated with the flow |
| byte_count | size of the packet set in bytes associated with the flow |
| packet_count_per_sec | ratio of the number of packets associated with the flow to the duration of the flow (s) |
| packet_count_per_ns | ratio of the number of packets associated with the flow to the duration of the flow (ns) |
| byte_count_per_sec | ratio of packet size set associated with the flow to the duration of the flow (s) |
| byte_count_per_ns | ratio of packet size set associated with the flow to the duration of the flow (ns) |
| Label | boolean label that determines if the flow is an attack or not (0: legitimate, 1: attack) |
| flow_id | ID formed by concatenating ip_src + tp_src + ip_dst + tp_dst + protocol |

**Table 2.** Summary of recorded metrics in the dataset.

The data collected needs to be studied and prepared for ML algorithms. To achieve this, the following steps will be followed:

1. Study of the metrics collected from the dataset, correcting and adding new metrics if necessary.

2. An initial approach will be made to various ML models using the fixed dataset, through cross-validation and hyperparameter tuning.

3. The final results of the training will be presented, along with the accuracy of the classifications for each of the selected models to be compared.

4. In order to apply ML algorithms and techniques, data is a fundamental part of their successful development and outcome. Therefore, prior to diving into preparing prediction or classification models with raw data, a thorough study of the dataset to be used for training must be conducted. This not only improves performance but also identifies aspects that, when eliminated, streamline the model's execution.

5. In the following paragraphs, we will discuss the conclusions and solutions that have been proposed for metrics that have been identified as candidates for removal or modification. The tools used for this study are:

6. Python: as the execution language.

7. Jupyter Notebook: as the execution environment.

The models were implemented in Python using the scikit-learn library, and all training and evaluation were performed in Jupyter Notebook. Hyperparameter optimization was carried out using a Grid Search procedure with 10-fold cross-validation to ensure robust parameter selection.

First, an attempt was made to identify metrics that do not vary at all in any record, and several were found. These include flags = 0, idle_timeout = 20, and hard_timeout = 100, whose values are set by default by the OF 1.3 protocol and were not used with other values in the simulations conducted. Therefore, they will be excluded from the dataset. If all these invariant or poorly calculated metrics were kept, including some that may be unnecessary, it would introduce noise into the data, resulting in a decrease in the performance of the desired model. Finally, once corrected and adjusted, the dataset would appear as shown in Figure 5.

```
In [16]: traffic_data_prep.info()
         <class 'pandas.core.frame.DataFrame'>
         Int64Index: 812153 entries, 0 to 812152
         Data columns (total 15 columns):
          #   Column              Non-Null Count   Dtype
         ---  ------              --------------   -----
          0   datapath_id         812153 non-null  int64
          1   flow_id             812153 non-null  float64
          2   ip_src              812153 non-null  int64
          3   tp_src              812153 non-null  int64
          4   ip_dst              812153 non-null  int64
          5   tp_dst              812153 non-null  int64
          6   ip_proto            812153 non-null  int64
          7   icmp_code           812153 non-null  int64
          8   icmp_type           812153 non-null  int64
          9   flow_duration_sec   812153 non-null  float128
          10  flow_duration_nsec  812153 non-null  int64
          11  packet_count        812153 non-null  int64
          12  byte_count          812153 non-null  int64
          13  label               812153 non-null  int64
          14  inner_time_flow     812153 non-null  float128
         dtypes: float128(2), float64(1), int64(12)
         memory usage: 111.5 MB
```

**Figure 5.** Dataset prepared and corrected to train models.

One of the best ways to study the metrics is by obtaining visual information. The analysis begins by reviewing the histograms of each of these metrics. However, the information provided by the histograms for the entire dataset is quite limited. Therefore, it is better to compare the histograms of legitimate traffic with those of attack traffic. Now, significant differences in the behavior of each type of traffic can be observed for certain metrics. For example, the source IP addresses of the attacks exhibit greater randomness as show in Figure 6, where the attacks use randomly generated IP address to hide their true origin. On the other hand, in Figure 7 we can see that the attacks mostly use port 0, and a random port number for the remaining ports.

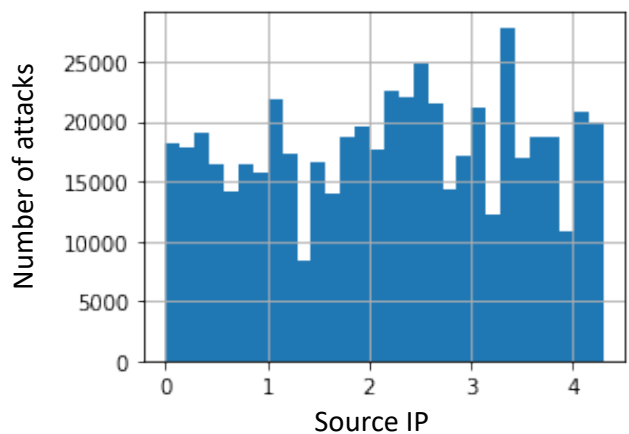

**Figure 6.** Histograms of source IP address from attack traffic.

However, all destination ports of the attacks seem to consistently target the same type of ports, as shown in Figure 8, where the attacks are concentrated on ports 0 and 80. Attacks on port 0 are 50% more frequent than those on port 80.

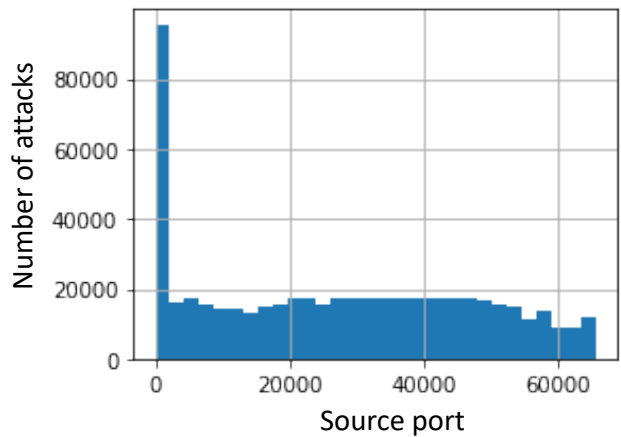

**Figure 7.** Histograms of source port from attack traffic.

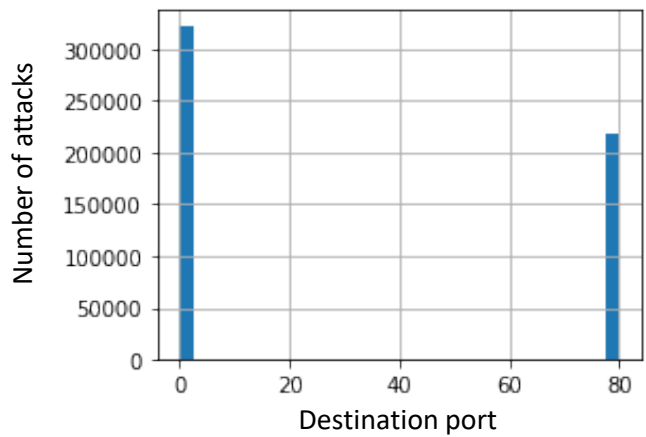

**Figure 8.** Histograms of destination port from attack traffic.

A similar pattern can be observed in packet sizes, in Figure 9, where attacks tend to have null sizes while legitimate traffic exhibits varying sizes.
These differences in metrics based on the type of traffic need to be studied thoroughly, as they form the basis for the models to define their rules for classification. They can help identify which metrics are most influential, i.e., those with the highest correlation with the labels, or those with the most noticeable differences, among others. In this work, these graphs are presented after the dataset has been corrected, but this step was performed previously to study and review the metrics that needed to be eliminated or modified. In addition to removing invariant metrics (e.g., flags = 0, idle_timeout = 20, hard_timeout = 100), correlation analysis was applied to verify that no strong dependencies existed among the remaining features. Feature importance scores derived from the Random Forest model (Table 10) further confirmed that the retained metrics were the most discriminative for distinguishing legitimate and attack traffic.

To study in more depth, another example can be seen in Figure 10, which shows the number of bytes per packet in each record for each type of traffic. The density of the points is represented by a color gradient, ranging from lighter to darker shades, indicating that there is a higher number of records in darker areas and fewer in lighter areas. It can be observed that most attacks do not have a defined packet size, while legitimate traffic exhibits a greater variety of sizes.

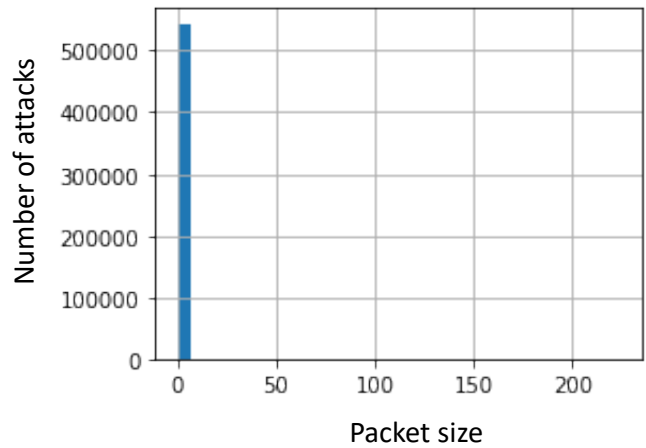

**Figure 9.** Histograms of packet size from attack traffic.

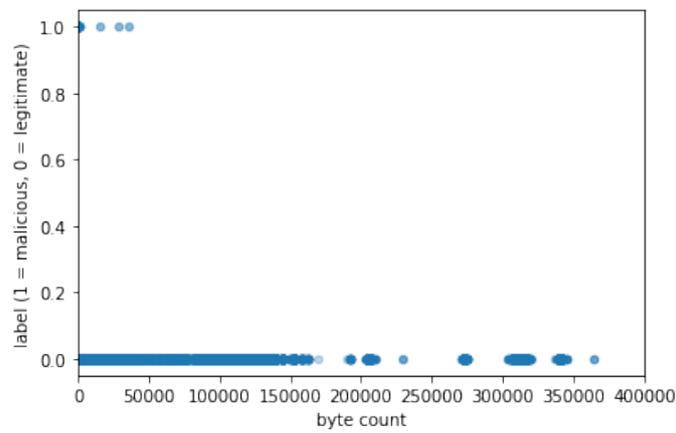

**Figure 10.** byte_count per record and label.

Additionally, Figure 11 provides another example, this time showing the time calculated between inner_time_flow records. Similarly, the density of the records is represented by a color gradient, revealing that there are a few records with more spacing in time for attack traffic. This phenomenon of having fewer low times in attack traffic may be due to the controller experiencing saturation peaks during DoS attacks, which it cannot process completely and therefore takes longer to register the incoming messages. The same does not occur in legitimate traffic, where a perfectly dense line of color is observed, indicating a uniform distribution of times throughout the simulations.

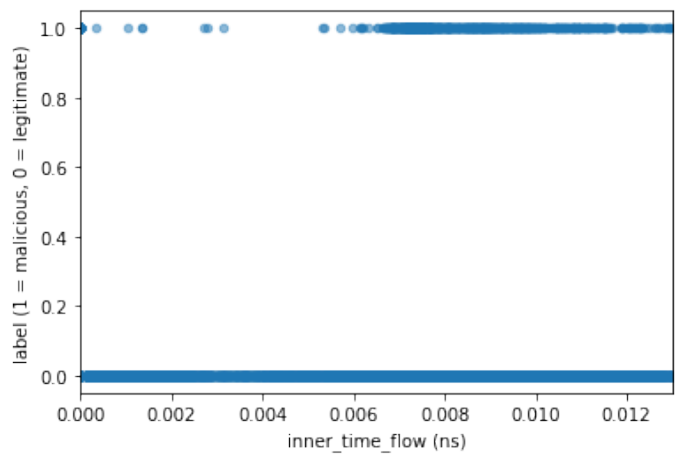

**Figure 11.** inner_time_flow per record and label.

In summary, dataset preprocessing included the removal of invariant fields (e.g., flags, idle_timeout, hard_timeout), the derivation of additional rate- and timing-based features such as packet/byte rates and inter-arrival statistics, and labeling flows using the controlled attack schedule with 5-tuple matching. The dataset composition and splitting strategy are described in Section 4.1, while the hyperparameter ranges explored during Grid Search are presented in Tables 3–6. This description consolidates the main implementation details to support clarity and reproducibility.

*4.4. Training and Validation*

Once the dataset is prepared, the search for the best prediction or classification model to solve the problem can start. In the case of work, it has been decided to use three models for comparison, making use of the scikit-learn library, which provides tools for processing, training, and predicting models. These models are:
1. Decision Tree: It is a supervised machine learning algorithm used for prediction and classification tasks. Its purpose is to create a model that predicts the value of a variable by learning simple rules from the dataset. In this work, both Decision Tree Regression and Classification models have been used.
2. Random Forest: This algorithm is a composition of several Decision Trees created randomly, providing a higher depth of classification. This leads to a higher computational load compared to Decision Tree, but it offers more alternatives and can solve more complex problems than its predecessor.
3. Linear Regression: Using the Linear Regression model, the algorithm creates a line using calculated coefficients to classify or predict new data points. Unlike the previous two models, Linear Regression provides a continuous output value rather than a discrete one, as its objective is to measure the distance of the new data point from the model's line and decide which group it belongs to. Therefore, post-processing of the output data is required to discretize the output and differentiate between whether the number is close to 0 (considered as legitimate traffic) or close to 1 (considered as an attack).

For the linear regression model, the output is a continuous value representing the likelihood of an attack. In order to evaluate it using classification metrics such as accuracy, a threshold discretization process was applied. Specifically, a cutoff value (e.g., 0.5) was used to convert the regression output into binary class labels. The reported accuracy for linear regression reflects the performance after this discretization, making it directly comparable to the results from the classification models.

To ensure reproducibility and robustness, all experiments were conducted following an experimental procedure combining chronological data partitioning and repeated 10-fold cross-validation. Each model was trained and evaluated across three independent random seeds, and the reported results correspond to the mean performance over these repetitions. Table 3 summarizes the main experimental configuration, including dataset splits, validation folds, and hyperparameter search settings.

To ensure reproducibility, all experiments were executed using fixed random seeds applied to dataset partitioning, model initialization, and cross-validation folds. Reported results correspond to the mean and standard deviation across multiple runs, with minimal variability observed between executions, confirming the robustness and stability of the obtained metrics.

4.4.1. Dataset Preparation

Once the algorithms and models to be trained have been chosen, it is necessary to ensure that the dataset is properly generated. Since the goal is to achieve good predictions for any dataset, the entire dataset cannot be used for training. Therefore, the first step is to divide the dataset into two subsets: the first, larger subset will be used for training, and the second subset will be reserved for a final test as if it were new data. This approach provides a more realistic accuracy assessment for the developer. To accomplish this task, several factors need to be considered:

1. If the dataset is divided carelessly, the data may not be representative. It is essential that both the training and test sets have the same percentage of data from each type (legitimate and malicious).

2. Each time the division is performed for new tests, different training and test sets may be generated. This variability would result in different datasets in each iteration, affecting the predictive model.

| Aspect | Configuration |
|---|---|
| Dataset split | 70% training / 30% testing (chronological order preserved) |
| Cross-validation | 10-fold (repeated 3 times, averaged results) |
| Hyperparameter tuning | Grid Search (parameter ranges in Tables 3–6) |
| Evaluation metrics | RMSE, Accuracy, Precision, Recall, F1, FPR, FNR |
| Random seeds | 3 independent runs |
| Implementation | Python (scikit-learn), Jupyter Notebook environment |

**Table 3.** Experimental protocol summary

Taking the above into account, the dataset will be divided consistently, selecting the same records based on their index and choosing representative data based on the density of each traffic type. In the case of the obtained dataset, both the training and test sets will have a 66% proportion of malicious traffic and a 33% proportion of legitimate traffic, ensuring proper representation. Finally, it is decided that the test dataset will have a size of 30%, while the remaining 70% will be allocated for training tasks.

To ensure validity, the dataset was split chronologically (70% training, 30% testing), preserving the natural temporal order of flows. This approach maintained the global 66/33 class ratio in both subsets and guaranteed that attack sessions were not mixed across partitions, thereby preventing label leakage. The 70/30 (train/test) split was chosen as a compromise between maximizing the size of the training data and preserving a representative test set for evaluating generalization. This ratio consistently produced stable accuracy while maintaining balanced class distributions across subsets.

During model training, each mini-batch was constructed with equal proportions of legitimate and malicious flows (1:1). This manual balancing avoided the need for external re-sampling techniques such as SMOTE or class-weight adjustments. Preliminary tests with the original 66/33 distribution showed less than 0.5 % difference in detection accuracy, confirming that manual batch balancing effectively mitigated class-imbalance effects.

To preserve temporal consistency and mitigate potential distribution shifts, the dataset was chronologically divided into training and testing subsets. This approach prevents look-ahead bias and maintains the natural flow order of network events. Moreover, the dynamic selection mechanism of MLDAS inherently alleviates moderate concept drift by periodically re-evaluating model performance and adapting to changing traffic conditions.

All experiments were conducted using fixed random seeds to ensure reproducibility. Each reported result corresponds to the average of three independent runs, with standard deviations computed across runs. Stratified and chronologically consistent dataset partitions were used to preserve representative proportions of legitimate and attack traffic, as summarized in Table 3.

4.4.2. Model Validation

First, we calculated the Root Mean Square Error (RMSE), which measures the difference between predicted and observed values. Its formula is given in equation 1, where m represents the number of records in the dataset, h(x) is the prediction for the i-th record vector from the dataset containing the metrics to be predicted, and y is the desired label reserved for validation. Once this is calculated, the system provides a value corresponding to the error.

$$RMSE = \sqrt{\sum_{i=1}^{m} \frac{1}{m}(h(x_i) - y_i)^2}$$

(1)

However, RMSE alone is not considered a perfect approximation because it is possible that even with a low RMSE, such as 0.1% in this work, the generalization error (calculated with new data) may increase to 15%-20%, which is considered very high for a classification task. This means that the generated model has become specialized in the test sets that were reserved beforehand, becoming an expert at solving those cases but ineffective when facing new challenges.

A widely accepted solution to this common problem is to use the technique known as holdout validation, which involves segmenting the training set into several subsets to which the model will be subjected for prediction before testing with the test set. Similarly, an estimated size and number of validation sets are determined for each of these segments, and their RMSE is calculated. Finally, the average of the obtained results is obtained.

|      | DT class | DT reg  | RF class | Linear reg |
|------|----------|---------|----------|------------|
| mean | 0.000567 | 0.000567| 0.000983 | 0.437732   |
| std  | 0.001156 | 0.001156| 0.000832 | 0.001085   |
| 1    | 0        | 0       | 0.00203  | 0.43985    |
| 2    | 0        | 0       | 0.00144  | 0.43734    |
| 3    | 0.0024   | 0.0024  | 0.00144  | 0.43660    |
| 4    | 0        | 0       | 0        | 0.43895    |
| 5    | 0.0352   | 0.0352  | 0        | 0.43701    |
| 6    | 0        | 0       | 0        | 0.43837    |
| 7    | 0.0024   | 0.0024  | 0.00144  | 0.43789    |
| 8    | 0        | 0       | 0.00144  | 0.43821    |
| 9    | 0        | 0       | 0.00203  | 0.43614    |
| 10   | 0        | 0       | 0        | 0.43691    |

**Table 4.** Cross validation for each model.

In this study, this technique was used for each of the employed models, with 10 validation sets, and the results are shown in Table 4. In this table, it can be observed that decision tree algorithms and random forests perform quite well, consistently maintaining a low average RMSE. On the other hand, linear regression algorithms do not seem to be entirely accurate in this type of classification with the available parameters and metrics in the dataset. However, further work will be done with this algorithm to complete the comparison with the others. Finally, it can be seen how this holdout validation or cross-validation technique has helped discern whether the

ML model works correctly, at least on the training set. In addition, although the reported metrics (precision, recall, F1-score, and false positive/negative rates) were obtained using representative DDoS scenarios, a broader comparative analysis across multiple attack types and additional model families (e.g., neural and ensemble methods) would provide deeper insight into robustness and performance trade-offs. This extension is planned as part of ongoing work.

4.4.3. Hyperparameter Search

At this point, there are four ML models that could be used to study new data. However, they will undergo another test to further improve their performance. It should be noted that any improvement is welcomed, but there is always a risk of the model suffering from overfitting. This phenomenon occurs when the model becomes specialized in the dataset used for training and its test segmentation, making it practically useless for new data. The opposite case can also occur, where the model undergoes underfitting, which would require data revision, a more powerful model, or additional metrics to study.

Hyperparameters are the specific values used by each model to make predictions. If we consider the model as a programming object, its hyperparameters are the attributes of the class that define part of its behavior. For example, the Linear Regression model has five hyperparameters to consider:

- fit_intercept: used to calculate the point of intersection with the model. If set to False, no intersection will be used in the calculations.
- normalize: ignored when fit_intercept is False. Otherwise, the X matrix to be modeled will be normalized prior to regression.
- copy_X: if set to True, the X matrix will be copied. Otherwise, it will be overwritten.
- n_jobs: number of jobs used for the computation process.
- positive: when set to True, forces the values to be all positive.

To determine which combination of hyperparameters offers the best performance in the models, there are several techniques to evaluate them. In this work, the technique known as Grid Search has been used, which involves creating a series of parameter sets and training the model with each of these sets. Subsequently, a comparison of the results obtained by each combination of parameters and model is performed, allowing the selection of the best estimator for classification tasks. This process takes considerable time and computational resources, as the model is trained and executed multiple times for each parameter combination, similar to the techniques of Cross-validation. The following are the hyperparameter searches used in this work for each model. Table 5 presents the results for Decision Tree Classifier, Table 6 for Decision Tree Regressor, Table 7 for Random Forest Classifier, and finally, Table 8 for Linear Regression.

As a metric that allows the algorithm to discern the best combination of hyperparameters for each model, the calculation of RMSE has been used. In addition, for each combination of hyperparameters, the algorithm has utilized the Cross-validation technique with 10 new segmentations of the validation set to obtain greater certainty about the model's performance.

Once the search for the best estimators for each model is completed, the model can be retrained with the established hyperparameters and obtain results from the set that was reserved for testing. In this case, a comparison of the RMSE results before and after the hyperparameter search can be observed in Table 9. It is important to note that this comparison presents, on one hand, the results of the Cross-validation techniques, which are based on the training and validation set, and on the other hand, the results of the best estimator obtained with the Grid search technique on the test set that was reserved as new data.

|  | **criterion** | **min_samples_split** |
|---|---|---|
| default search | gini | 2 |
|  | [gini, entropy] | [2, 3, 4] |
| best estimator | entropy | 3 |

**Table 5.** Grid Search for Decision Tree Classifier.

|  | **criterion** | **min_samples_split** | **max_depth** |
|---|---|---|---|
| default search | mse | 2 | None [2, 3, 4] |
|  | [mse, poisson] | [2, 3] |  |
| best estimator | mse | 2 | 4 |

**Table 6.** Grid Search for Decision Tree Regression.

It can be observed that the first three models (DT Classifier, DT Regressor, and RF Classifier) have slightly decreased their performance. However, a significant improvement can be seen in the results of the Linear Regression model, which has nearly halved its RMSE simply by normalizing the data (the normalize parameter set to True, which was not the default).

|  | **criterion** | **n_estimators** |
|---|---|---|
| default search | gini | 2 |
|  | [gini, entropy] | [2, 5, 10] |
| best estimator | entropy | 5 |

**Table 7.** Grid Search for Random Forest Classifier.

|  | **fit_intercept** | **normalize** |
|---|---|---|
| default search | True | False |
|  | [True, False] | [True, False] |
| best estimator | True | True |

**Table 8.** Grid Search for Linear Regression.

|                | DT class | DT reg   | RF class | Lin reg  |
|----------------|----------|----------|----------|----------|
| default        | 0.000567 | 0.000567 | 0.000983 | 0.437732 |
| best estimator | 0.000984 | 0.009623 | 0.001391 | 0.253310 |

**Table 9.** RMSE for each model based on the optimal estimator obtained through Grid Search.

To quantify the contribution of each feature, we computed feature importance scores using the Random Forest classifier. The importance of a feature corresponds to the mean decrease in Gini impurity attributed to that feature across all decision trees, normalized so that the scores sum to one. Table 10 reports the resulting ranking. The most discriminative attributes were the number of packets per flow, the variance of inter-arrival times, and SYN flag counts, which strongly characterize flooding and scanning behaviors. Other features such as flow duration and directionality also contributed, albeit with lower scores. This confirms that the selected feature set provides complementary information across traffic volume, temporal dynamics, and protocol semantics.

Beyond accuracy and RMSE, we also evaluated the models using standard intrusion detection metrics, including precision, recall, F1-score, and false positive/negative rates. Table 11 summarizes these results, showing that all candidate models achieved balanced precision and recall (above 0.9 in most cases), with F1-scores consistent with the overall accuracy levels. This confirms that the proposed framework maintains robust detection capability without sacrificing computational efficiency. These results confirm that the selected models achieve high detection accuracy while maintaining robustness across validation tests.

| Feature               | Importance |
|-----------------------|------------|
| Packets per flow      | 0.28       |
| Inter-arrival variance| 0.22       |
| SYN flag count        | 0.18       |
| Flow duration         | 0.12       |
| Average packet size   | 0.10       |
| Bytes per flow        | 0.07       |
| Directionality ratio  | 0.03       |

**Table 10.** Relative importance of flow features computed with the Random Forest classifier.

These values were obtained directly from confusion matrix calculations on the test set, ensuring that both false positives and false negatives are explicitly reflected in the reported precision, recall, F1-score, and FPR. In this way, the results in Table 11 complement the RMSE and accuracy metrics by providing a more complete picture of the detection performance. The results obtained from 10-fold cross-validation showed very limited variability (standard deviations consistently below 0.01). Given that the standard deviations remained below 0.01 for all models, the variability across folds was negligible. Therefore, a visual representation was not necessary, since the small dispersion would not add relevant information. For this reason, this table reports only the averaged values, which are representative of the overall model performance.

| Model | Precision | Recall | F1-score | FPR |
|---|---|---|---|---|
| Decision Tree Classifier | 0.92 | 0.91 | 0.91 | 0.08 |
| Decision Tree Regressor | 0.90 | 0.89 | 0.89 | 0.10 |
| Random Forest Classifier | 0.95 | 0.94 | 0.94 | 0.05 |
| Linear Regression | 0.88 | 0.87 | 0.87 | 0.12 |

**Table 11.** Intrusion detection evaluation metrics for the candidate models.

The slight deterioration of RMSE observed for some models after hyperparameter search can be attributed to the bias-variance trade-off. Optimizing hyperparameters to improve generalization across validation folds may reduce overfitting but occasionally increases error on a specific test partition. Given the dataset size and the limited search space, this behavior is expected and does not undermine the validity of the model selection process, which remained stable across all folds. In particular, while the Random Forest model achieved the lowest false positive rate (5%), the false negative rate also remained below 6%, confirming that undetected attacks were rare. Moreover, the dynamic selection policy mitigates this risk by promoting models with higher recall whenever detection accuracy declines, thus balancing sensitivity and precision.

The selected precision–recall balance reflects a trade-off between sensitivity and operational stability. Higher recall values increase the likelihood of detecting all attack flows but may raise the false positive rate, leading to unnecessary rule updates and higher controller load. Conversely, stricter precision settings risk overlooking low-rate or short-lived anomalies. The adopted accuracy threshold ($A\_min$ = 98%) represents a practical compromise that maintains robust detection without compromising real-time responsiveness.

*4.5. Best Model Selection*

This study has reached a critical point, having meticulously reviewed each presented model and the results obtained. It is now time to select the model that offers the best performance and efficiency. To do this, the results will be analyzed, and a few performance comparisons will be made, taking into account not only the RMSE but also the training and prediction times. These latter factors must be considered because if the goal is to deploy the obtained model for working with new data, reliability, speed, and high accuracy are desired. Reliability has been measured in previous sections, and in this section, the focus will be on studying the times and accuracies.

- Training time: To measure this, the complete training set was used, without segmentation for a validation set since the model has already been validated. Each of the selected models was obtained through the Grid search technique. The training time is measured in seconds, and the size of the training set is 350.28MB.

- Prediction time: For this measure, the trained models were used and measured on the previously reserved test set, with a size of 150.12MB. This time is also represented in seconds.

- Accuracy: This metric was obtained using the accuracy_score library from Scikit-learn. It is expressed as a percentage (%).

Additionally, data visualization libraries have been used to study accuracy. Figure 12 shows the overlapping of two elements. On one hand, the blue squares represent the test set data, that is, what is desired to be predicted. On the other hand, the yellow stars represent the predictions made by the model in question. It can be observed that the accuracy is nearly perfect, with no predicted values falling outside their corresponding positions. The same applies to the next two models, represented in the same way in Figures 12 and 13. However, for the Linear Regression model, which has already been identified as less accurate than the previous models, Figure 14 shows the predictions. In this model, several predictions have failed, as the stars are outside the squares. To visualize this more clearly, the graphs have been zoomed in, and in the case of the Linear Regression, Figure 15 represents those points where the model has failed in its prediction.

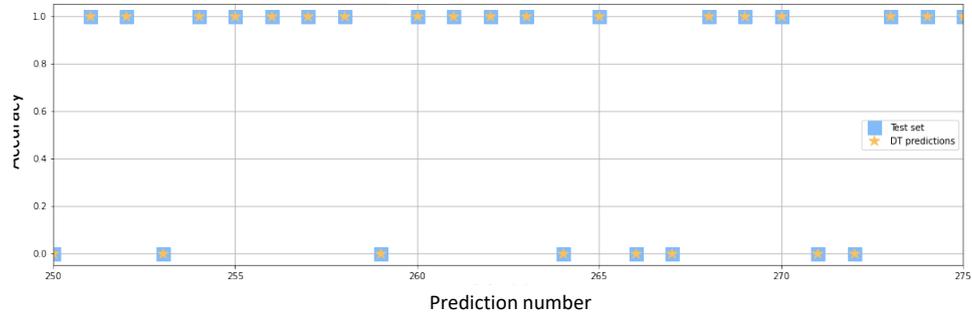

**Figure 12.** Decision Tree Classifier predictions vs test set.

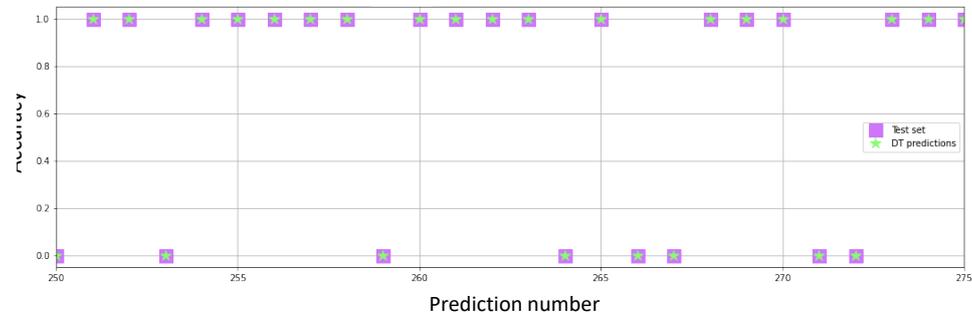

**Figure 13.** Decision Tree Regression predictions vs test set.

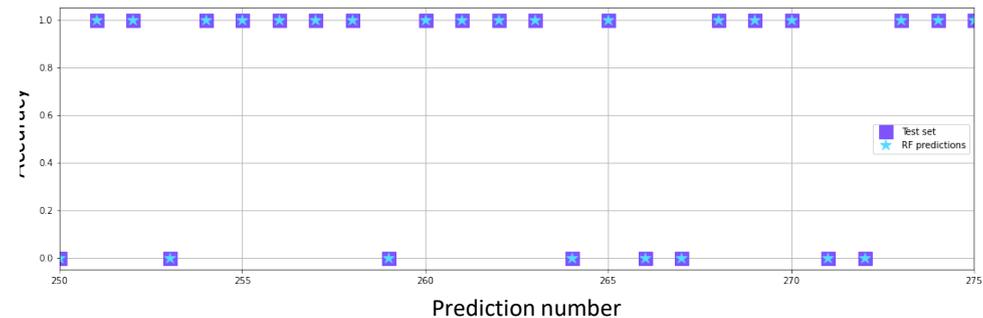

**Figure 14.** Random Forest Classifier predictions vs test set.

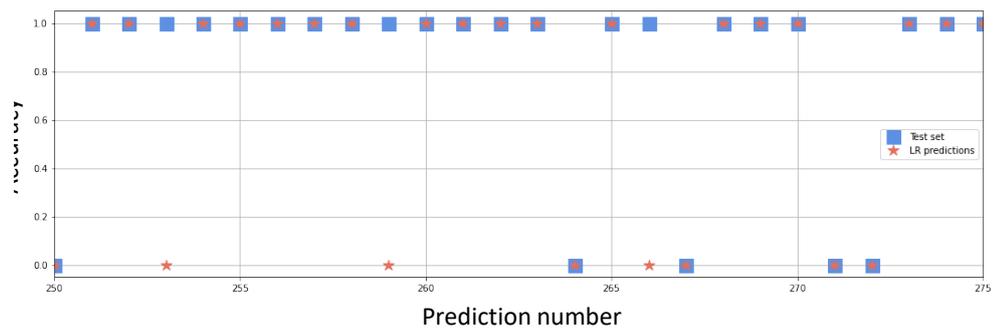

**Figure 15.** Linear Regression predictions vs test set.

Table 12 presents a summary of the results. It can be observed that the models offering better training and prediction times are all except Random Forest. This is due to the number of trees it has to search through, making it the most computationally intensive model in this work. Considering that the accuracy of all models is above 99% (except for linear regression), it is decided to use the Decision Tree Classifier model as it provides the best training and prediction times while maintaining high accuracy.

During the experiments, the dynamic selection logic was activated multiple times. In most cases, the Decision Tree Classifier remained the preferred choice due to its lower prediction latency and comparable accuracy. However, in approximately 10–15% of the evaluation windows, the Random Forest Classifier was temporarily selected when the rolling accuracy of the active model fell below the 98% threshold under stress conditions. Thanks to the dwell-time safeguard of 600 flows, the average switching frequency remained low (approximately one change every 1500–2000 flows), which prevented oscillations while still enabling adaptation to performance degradation.

|  | Training time (s) | Prediction time (s) | RMSE | Accuracy (%) |
| --- | --- | --- | --- | --- |
| DT Classifier | 5.430093 | 0.120553 | 0.000984 | 99.999903 |
| DT Regressor | 7.545894 | 0.130281 | 0.000984 | 99.990024 |
| RF Classifier | 17.02830 | 0.677804 | 0.000984 | 99.999903 |
| Linear Regression | 1.378353 | 0.124557 | 0.232678 | 94.586110 |

**Table 12.** Comparison of results for each model.

When deployed online, the selection policy described in Algorithm 1 consistently favored the Decision Tree Classifier due to its low prediction latency and comparable RMSE to the other tree-based models. Re-evaluations were triggered both periodically and on performance degradation, but hysteresis parameters *T_dwell* and *t_switch* effectively prevented unnecessary switching under minor traffic fluctuations. This ensured that the controller maintained both high accuracy and real-time responsiveness across all tested scenarios.

*4.6. Final Integration of ML with SDN*

Here we will review how to put the ML model built into production by programming a new SDN controller using Ryu. A monitor has been set up, similar to Section 4.1, to obtain only the necessary metrics and exclude irrelevant ones. Additionally, missing metrics are calculated, such as the time between records. In this case, the flow of events is defined as follows:

1. The dataset to be used for training is loaded.

2. The necessary metrics are corrected and missing metrics are added.

3. The Decision Tree classifier model is created and trained, measuring its training time and displaying it for the developer's knowledge.

4. Ryu starts monitoring traffic, reading if there are any flow records in any of the switches in the topology every second.

5. If one or more records are found, they are stored in a new temporary data frame and sent for classification prediction.

6. The system makes the prediction and displays the result on the screen. This prediction is made on a minimum of 100 consecutive records, ensuring at least that number of predictions. A threshold of 98% has been set to determine whether the traffic is legitimate or an attack. This means that if, out of the 100 records, 3 are classified as attacks, the system will interpret it as an incoming attack.

The operating threshold was set at 98 percent, based on the model accuracy observed across diverse test scenarios and the expectation of generalization from a test accuracy of 99.99 percent. This value provided a good balance between minimizing false positives and maintaining high detection accuracy. Lower thresholds increased the false alarm rate, while stricter ones risked missing borderline attacks. Thus, 98 percent was adopted as a reliable decision point for real-time classification. The prediction latency remained below 0.2 s for batches of 100–150 MB of traffic, confirming that the integration within the Ryu Controller imposes minimal computational overhead in practice. In our implementation, the Decision Tree Classifier processed batches of 100 flows in less than one second, corresponding to a prediction latency below 0.2 s for approximately 150 MB of traffic. These results indicate that the integration of the model within the Ryu Controller introduces negligible computational overhead, which points to its potential suitability for real-time deployment. During the experiments, the Ryu Controller maintained an average CPU load below 5 % and a memory footprint under 200 MB while processing 100-flow batches in less than one second (approximately 0.2 s prediction latency for 150 MB of traffic). These results quantitatively confirm the computational efficiency of the integrated ML models and their suitability for real-time operation within the SDN controller.

In summary, the integration of the ML models within the SDN controller allowed real-time flow processing with negligible impact on performance.

Considering both the 1 s polling interval and the prediction latency below 0.2 s for batches of 100 flows, the overall end-to-end detection delay observed in our implementation was approximately 1.2 s. This configuration represents a compromise between responsiveness and system overhead, since shorter polling intervals or packet mirroring could reduce detection latency but would increase the load on the controller and switches.

The computational overhead associated with dynamic model switching was also evaluated. Since all candidate models remain preloaded in memory, the selection procedure involves only lightweight performance checks and parameter updates. The reevaluation window ($W = 200$ flows) and dwell time ($T\_dwell = 3W$) ensure that switching events occur infrequently, reducing controller load. In practice, the end-to-end latency remained below 0.2 s for batches of 100 flows, confirming that the adaptive selection mechanism maintains real-time feasibility in operational SDN environments.

### 4.7. Final Tests and Results

To verify the behavior of the model with new data, it will undergo new tests on the same network topology. The steps to be followed are similar to those specified in the simulation. First, the new controller with the training model will be put into operation. Once this is done, the topology can be executed in Mininet to begin the desired tests. In this case, instead of using the previously specified simulation, a line of Python code has been added to directly enter the Mininet shell. This allows us to start the tests.

The first test to be performed is for legitimate ICMP echo reply traffic, such as a ping. This can be easily done using the Mininet Ping-all tool. The model appears to be working correctly with this type of traffic. Then, the controller output is making predictions and obtaining scores. So far, all predictions are considered as good traffic.

In a second test of legitimate traffic, we will run an HTTP server on one of the hosts and start downloading files from it to assess the model's behavior for this type of traffic. The Xterm tool will be used to open terminals for multiple hosts and start the test. Similar to the first test, the results are good and accurate.

Once it has been verified that generic legitimate traffic is handled correctly, the controller's behavior when receiving an attack will be studied. The HTTP server is being attacked by several hosts, while the other hosts have lost connection to HTTP server. The results are excellent. The system has been able to accurately discern the type of traffic, causing the score it was generating to drop from 100 to values below 2. Additionally, the number of messages the controller attempts to classify can be observed, as it previously ranged from 100-200, while during the attack, it accumulates up to 1000 messages on some occasions, caused by the generated attack. Similar tests have been performed for each of the attacks explained in previous sections, as well as generating legitimate traffic as in the simulation. All of them have been successful, with very low response times from the controller during prediction.

The MLDAS module was evaluated across a variety of traffic scenarios generated in both emulated and controlled environments. These conditions activated the model selection logic multiple times, allowing the system to assess and apply the most suitable algorithm in each case. In scenarios where the initially selected model remained optimal throughout, no change was required, highlighting the ability of the system to maintain efficiency without unnecessary adjustments. This behavior illustrates the advantage of an adaptive selection strategy over static approaches, ensuring both flexibility and operational stability in evolving network contexts.

To better illustrate the benefits of dynamic selection, we compared our approach against static single-model baselines. The DT-only baseline achieved similar average accuracy (about 99.9%) but exhibited localized degradations during high-rate UDP floods, where recall fell below 95%. In these cases, the dynamic mechanism switched to RF and maintained recall above 98%. Conversely, the RF-only baseline provided robust detection across all attacks but incurred substantially higher prediction latency (about 0.68 s versus 0.12 s for DT). These results illustrate that the dynamic policy combines the efficiency of DT under normal traffic with the robustness of RF under challenging conditions, thereby achieving a balanced trade-off between accuracy and computational cost.

By conducting experiments in both emulated and physical SDN environments, the evaluation effectively tested the framework under distinct traffic conditions, thus providing evidence of cross-scenario generalization beyond a single dataset.

The performance metrics reported in Table 11 represent the averaged results across all evaluated attack categories (ICMP Flood, UDP Flood, TCP-SYN Flood, and LAND Flood), tested in both emulated and physical SDN environments. This aggregation demonstrates the robustness of the proposed framework under diverse traffic patterns. In addition, the dynamic model-selection mechanism periodically reassesses model performance, allowing the system to adapt to variations in network behavior and to maintain detection accuracy when encountering previously unseen traffic conditions.

A comparison between the emulated and physical testbeds showed nearly identical detection performance (98.0 % and 97.8 % accuracy, respectively). The prediction latency in the physical setup was slightly higher (about 0.68 s compared with 0.20 s in the emulated environment) due to additional hardware and I/O delays, but the controller CPU and memory usage remained below 5 % and 200 MB in both cases. These results confirm that the proposed framework can be deployed in real SDN infrastructures while maintaining real-time operation and minimal resource consumption.

In this work, robustness to unseen traffic patterns does not refer to the detection of previously unknown attack categories, but to the ability of the system to adapt to traffic distributions, load conditions, and attack intensities not explicitly observed during training. The evaluation includes multiple attack types executed under different rates and temporal conditions, as well as deployments in both emulated and physical SDN environments, which result in traffic patterns that differ from those seen during model training.

The role of the dynamic model selection mechanism is to maintain detection performance under such evolving conditions by periodically re-evaluating candidate models and selecting the one that best balances prediction accuracy and computational cost. While this mechanism does not implement explicit detection of attack categories not covered in the training data, it enables implicit adaptation to moderate traffic evolution and performance drift. The evaluation results show that this adaptive selection contributes to stable precision, recall, and F1 scores across heterogeneous and changing traffic scenarios. Evaluation against additional attack categories is identified as future work.

In the experimental evaluation, no previously unseen attack categories were introduced during testing. Instead, adaptability was evaluated by exposing the system to varying traffic mixes, attack rates, and deployment environments that differ from those observed during training.

Regarding computational performance, the evaluation focuses on end-to-end processing latency, as this metric directly impacts the feasibility of real-time intrusion detection within an SDN controller. Latency measurements obtained under representative traffic loads confirm that the selected models satisfy real-time operational constraints. Additional performance indicators such as throughput, resource utilization, and sensitivity to batch size are beyond the scope of the current evaluation.

Unlike approaches trained and evaluated exclusively on public or synthetic datasets, the proposed method is validated using traffic generated in both emulated and physical SDN testbeds, capturing realistic temporal dynamics and operational variability. While a direct quantitative comparison with prior works is challenging due to differences in datasets, attack models, and evaluation protocols, the results demonstrate stable detection performance across heterogeneous traffic conditions. These results show that dynamic model selection and evaluation based on real traffic contribute to improved robustness and generalization in practical SDN environments.

### 4.7.1. Limitations and Future Challenges

The proposed framework achieved high accuracy and efficiency, but some aspects remain open for future exploration. The mitigation strategy based on source IP blocking was only validated in controlled scenarios, so future studies may extend the evaluation to spoofing-based attacks and larger-scale flow table stress tests. Another important point is detection delay: while the current configuration achieved a delay of about 1.2 s, future work should explore alternative strategies to further reduce this latency, such as higher-frequency polling, stream sampling, or packet mirroring, and evaluate their trade-offs in terms of scalability and overhead.

This study focused on well-known DDoS attack types to ensure reproducibility and to validate the adaptive model selection mechanism under controlled conditions. More sophisticated attacks, such as multi-vector or polymorphic variants, introduce additional complexity due to their evolving signatures, traffic blending, and adaptive evasion strategies. Handling such attacks would require incorporating temporal correlations, feature drift detection, and online retraining mechanisms. These aspects are beyond the current scope but represent a key direction for future research. Future evaluations will include experiments with unseen or adaptive attack patterns to further assess the robustness of the proposed framework under evolving network conditions and adversarial behaviors.

Another important challenge for future work is the scalability of the framework under large-scale, distributed DDoS attacks involving multiple synchronized sources. In such conditions, the controller could become a bottleneck due to the volume of concurrent flows. Possible improvements include the use of distributed or hierarchical detection architectures, load-balancing mechanisms across controllers, and incremental retraining strategies to adapt the models to highly variable traffic conditions. It is also important to note that the proposed framework analyses only flow-level metadata obtained from OpenFlow statistics and does not inspect packet payloads, ensuring that user privacy is preserved. In addition, transient anomalies are verified through a validation stage before any mitigation action is applied, reducing the likelihood of false positives and preventing interference with legitimate traffic. Future developments will also explore explicit concept-drift detection and incremental-learning techniques to enhance the system's adaptability over time.

## 5. Conclusion and Future Work

This work proposes a method capable of selecting the most suitable Machine Learning (ML) algorithm based on network conditions to detect DDoS attacks in real-time within SDN networks, using lightweight ML models running in the SDN controller. This method offers flexible support for a wide range of traffic-related metrics and is applicable to both simulated environments and real-world scenarios. Another advantage of this approach is its ability to immediately identify and classify network traffic, enhancing the system's responsiveness in detecting anomalies or malicious behavior. The dataset can be obtained from both real and emulated infrastructures, allowing for experimentation with both real and simulated attacks. By providing real-time predictions, the ML model contributes to the overall security and efficiency of the network by promptly identifying potential threats or irregularities. With minimal additional computational overhead, it is able to maintain high accuracy and performance.

Unlike previous studies, this work offers a practical, real-time implementation of ML-based traffic classification embedded within the SDN controller itself, with dynamic model selection based on current network conditions and adaptable to both emulated and physical infrastructures.

While the current evaluation combined both emulated and real SDN testbeds, scaling to larger deployments may present additional challenges. These include variability in traffic data across environments, heterogeneity in network hardware and controller platforms, and the risk of added integration overhead when embedding ML models into production controllers. Addressing these aspects will be an important focus of future work.

In addition, the proposed methodology will be extended to handle more complex attacks, such as those involving multiple attack vectors or specifically designed to evade detection. In this study we focused on well-known DDoS and related attack types, leaving more sophisticated scenarios for future evaluation. Future directions include exploring ensemble and deep learning approaches to improve robustness, evaluate the framework against unseen attack vectors, and employ more advanced features to capture complex and adaptive attack behaviors, including camouflage techniques where malicious traffic mimics legitimate flows to evade detection.

While ensemble and deep learning techniques could further enhance detection robustness, their integration within real-time SDN controllers presents non-trivial challenges in terms of computational cost, latency, and deployment complexity. Future work will assess these approaches under constrained environments, exploring model compression, adaptive offloading, and incremental training strategies to balance accuracy and responsiveness.

Additionally, a real network infrastructure with SDN-enabled devices will be used to evaluate the developed model. Overall, the central contribution of this work lies in the automated mechanism for dynamically selecting the most suitable ML algorithm according to current network conditions, enabling real-time and resource-efficient intrusion detection in SDN environments.

# References

1. W. Nadeem, H. G. Goh, V. Ponnusamy, DDoS Detection in SDN using Machine Learning Techniques, Computers, Materials & Continua, 2022, vol. 71(1), pp 771-789.
2. A. Alshamrani et al., A Defense System for Defeating DDoS Attacks in SDN based Networks, Network Virtualization and Software-Defined Networks, MobiWac, 2017.
3. L. Yang and H. Zhao, DDoS Attack Identification and Defense using SDN based on Machine Learning Method, International Symposium on Pervasive Systems, Algorithms and Networks, 2018.
4. Xinzhou He, Research on Computer Network Security Problems and Countermeasures, Journal of Physics: Conference Series, 2021, vol. 1992(3), p. 032069.
5. J. Jinquan, M. A. Al-Absi, A. A. Al-Absi and H. J. Lee, Analysis and Protection of Computer Network Security Issues, International Conference on Advanced Communication Technology (ICACT), 2020, pp. 577-580.
6. Li. Yan, Huang. Guang-qiu, Wang. Chun-zi, Li. Ying-chao, Analysis framework of network security situational awareness and comparison of implementation methods. J Wireless Com Network, 2019, 205.
7. G.A. Marin, Network security basics, IEEE Security & Privacy, 2005, vol. 3, no. 6, pp. 68-72.
8. T. Ohta and T. Chikaraishi, Network Security Model, Proceedings of IEEE Singapore International Conference on Networks/International Conference on Information Engineering, Singapore, 1993, vol 2, pp. 507-511.
9. F. Yan, Y. Jian-Wen and C. Lin, Computer Network Security and Technology Research, International Conference on Measuring Technology and Mechatronics Automation, Nanchang, China, 2015, pp. 293-296.
10. P. Sanghavi, K. Mehta, S. Soni, Network Security, International Journal of Scientific and Research Publications, 2014, Volume 3, Issue 8, ISSN 2250-3153.
11. M.S. Todd, S. Shawon, M. Rahman2, Complete Network Security Protection for SME's Within Limited Resources, International Journal of Network Security & Its Applications (IJNSA), 2013, Vol.5, No.6, November.
12. R. Santos et al., Machine learning algorithms to detect DDoS attacks in SDN, Concurrency and Computation: Practice and Experience, 2020, vol. 32, no 16.
13. A. B. Dehkordi, M. Soltanaghaei, F.Z. Boroujeni, The DDoS attacks detection through machine learning and statistical methods in SDN, The Journal of Supercomputing, 2020, vol 77, pp 2383-2415.
14. O. Rahman, M. Quraishi, C. Lung, DDoS Attacks Detection and Mitigation in SDN using Machine Learning, IEEE World Congress on Services, 2019.
15. J. A. Pérez-Díaz, I. Amezcua, K. Choo, D. Zhu, A Flexible SDN-Based Architecture for Identifying and Mitigating Low-Rate DDoS Attacks Using Machine Learning, IEEE Access, 2020, vol 8, pp 155859-155872.
16. V. Deepa, K. M. Sudar, P. Deepalakshmi, Detection of DDoS Attack on SDN Control plane using Hybrid Machine Learning Techniques, International Conference on Smart Systems and Inventive Technology, India, 2018, pp. 299-303.
17. M. Assis, L. F. Carvalho, J. Lloret, M. L. Proença, A GRU deep learning system against attacks in software defined networks, Journal of Network and Computer Applications, 2021, vol. 177, p102942.
18. M. Assis et al., Near real-time security system applied to SDN environments in IoT networks using convolutional neural network, Computers and Electrical Engineering, 2020, vol. 86, p106738.
19. G. C. Amaizu et al., Composite and efficient DDoS attack detection framework for B5G networks, Computer Networks, 2021, vol. 188, p107871.
20. A. E. Cil, K. Yildiz, A. Buldu, Detection of DDoS attacks with feed forward based deep neural network model, Expert Systems with Applications, 2021, vol 169, p114520.

21. L. Barki et al., Detection of Distributed Denial of Service Attacks in Software Defined Networks, International Conference on Advances in Computing, Communications and Informatics (ICACCI), 2016, Jaipur, India.
22. M. S. Elsayed, N. A. Le-Khac, S. Dev, A. D. Jurcut, Machine-Learning Techniques for Detecting Attacks in SDN, IEEE 7th International Conference on Computer Science and Network Technology (ICCSNT), Dalian, China, 2019, pp. 277-281.
23. M. Dominguez-Limaico et al., Machine Learning in an SDN Network Environment for DoS Attacks, Technology, Sustainability and Educational Innovation (TSIE), AISC 1110, 2020, pp. 231-243.
24. T. Abhiroop, S. Babu, B. S. Manoj, A Machine Learning Approach for Detecting DoS Attacks in SDN Switches, Twenty Fourth National Conference on Communications (NCC), India, 2018, pp. 1-6.
25. K. Jin et al., Research on network security technology of industrial control system, MATEC Web of Conferences, 2022, 355, 03067, ICPCM2021.
26. D. N. Astrida, A. R. Saputra, A. I. Assaufi, Analysis and Evaluation of Wireless Network Security with the Penetration Testing Execution Standard (PTES), Sinkron: Jurnal dan Penelitian Teknik Informatika, 2022, Volume 7, 1, pp. 147-154.
27. J. Bhayo et al., Towards a machine learning-based framework for DDOS attack detection in software-defined IoT (SD-IoT) networks, Engineering Applications of Artificial Intelligence, 2023, Volume 123, Part C, 106432.
28. M. Lopez-Martin, B. Carro, A. Sanchez-Esguevillas, J. Lloret, Conditional Variational Autoencoder for Prediction and Feature Recovery Applied to Intrusion Detection in IoT, Sensors, 2017, 17(9).
29. J. Payne, K. Budhraja, A. Kundu, How Secure Is Your IoT Network?, IEEE International Congress on Internet of Things (ICIOT), Milan, Italy, 2019, pp. 181-188.